\documentclass[11pt]{article}

\usepackage{epsf}
\usepackage{amsmath}
\usepackage{amssymb}
\usepackage{graphicx}
\usepackage{dcolumn}
\usepackage{bm}

\begin{document}

\def\thefootnote{\fnsymbol{footnote}}

\vspace*{-1.5cm}
\begin{flushright}
\tt MTA-PHYS-0701
\end{flushright}

\vspace{0.2cm}
\begin{center}
\Large\bf\boldmath
One-loop quantum corrections to cosmological scalar field potentials
\unboldmath
\end{center}

\vspace{0.4cm}
\begin{center}
A. Arbey\footnote{Electronic address: \tt arbey@obs.univ-lyon1.fr}\\[0.4cm]
{\sl Universit\'e de Lyon, Lyon, F-69000, France ; Universit\'e Lyon~1,
Villeurbanne, F-69622, France ; Centre de Recherche Astrophysique de
Lyon, Observatoire de Lyon, 9 avenue Charles Andr\'e, Saint-Genis Laval cedex, F-69561, France ; CNRS, UMR 5574 ;\\ Ecole Normale Sup\'erieure de Lyon, Lyon, France}\\[0.8cm]
F. Mahmoudi\footnote{Electronic address: \tt nmahmoudi@mta.ca}\\[0.4cm]
{\sl Department of Physics, Mount Allison University, 67 York Street, Sackville,\\
 New Brunswick, Canada E4L 1E6}
\end{center}
\vspace{0.5cm}

\begin{abstract}
\noindent We study the loop corrections to potentials of complex or coupled real scalar fields used in cosmology to account for dark energy, dark matter or dark fluid. We show that the SUGRA quintessence and dark matter scalar field potentials are stable against the quantum fluctuations, and we propose solutions to the instability of the potentials of coupled quintessence and dark fluid scalar fields. We also find that a coupling to fermions is very restricted, unless this coupling has a structure which already exists in the scalar field potential or which can be compensated by higher order corrections. Finally, we study the influence of the curvature and kinetic term corrections.
\end{abstract}
\vspace{0.3cm}

\section{Introduction}
\begin{figure}[t]
\hspace*{0.5cm}\includegraphics[height=4cm]{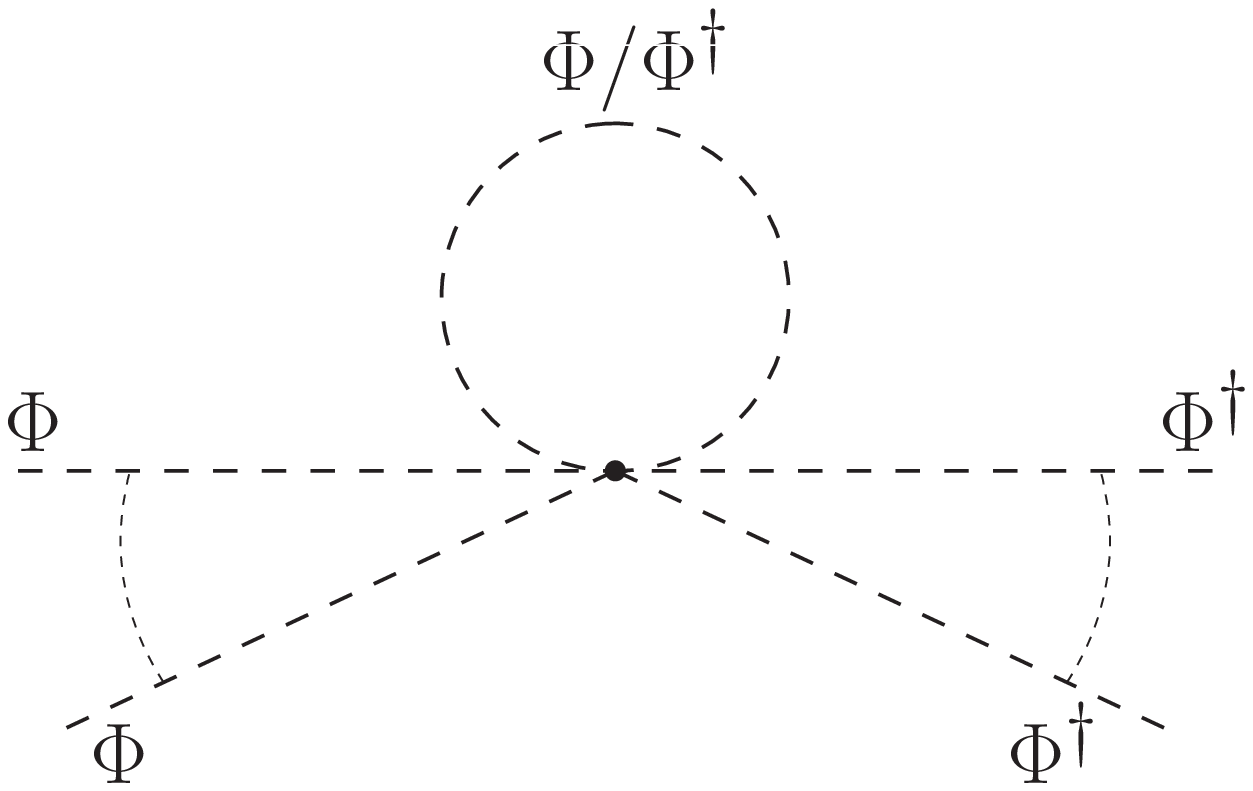}\\
\vspace*{-3.1cm}~\\
\hspace*{8.5cm}\includegraphics[height=2cm]{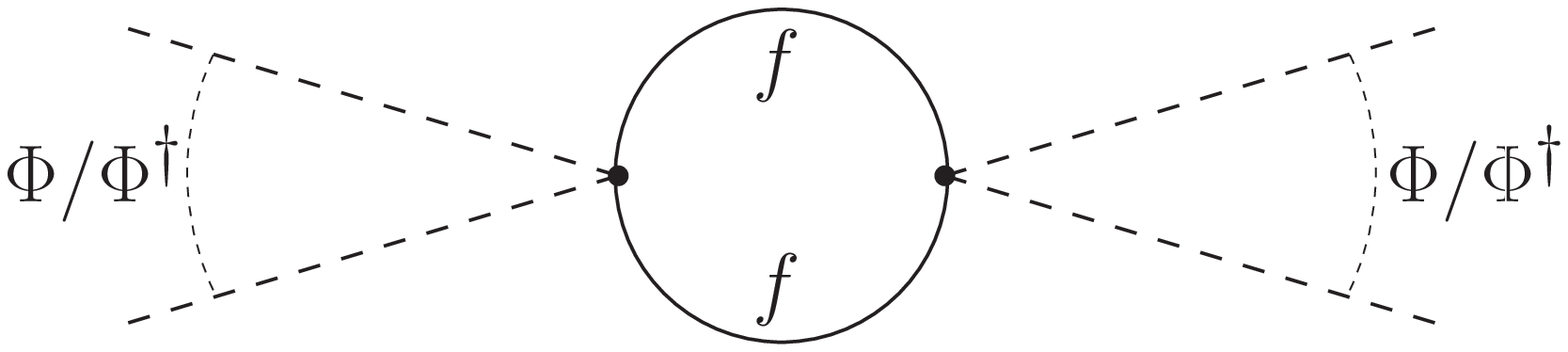}\\
\hspace*{3.5cm}(a)\hspace*{8.4cm}(b)
\caption{One-loop corrections to the potential. The dashed lines describe scalar fields, and the plain lines describe fermions. Diagram (a) represents the scalar corrections, whereas diagram (b) depicts the fermionic corrections. The multiple external lines correspond to arbitrary powers of $\Phi$ and $\Phi^\dagger$ in the potential.}
\label{fig}
\end{figure}

\noindent Dark energy and dark matter appear as two of the most important questions of modern cosmology. Many models trying to answer these questions involve scalar fields. In particular, to explain the behavior of dark energy, quintessence models \cite{quint, brax1999, bento2001} consider scalar fields with potentials dominating today, giving birth to a negative pressure. But dark matter can also be explained through scalar fields, as in boson star models \cite{bosonstar} or scalar field dark matter models \cite{scaDM, arbey_quadratic, arbey_cosmo, arbey_quartic}. A unification of dark matter and dark energy can also be performed into a dark fluid \cite{darkfluid1}, which can again be modeled with a dark fluid scalar field \cite{arbey_darkfluid2}. Thus, scalar fields are involved in many cosmological models.\\
One of the main difficulties to build scalar field-based models is the choice of a potential. The usual way to deal with scalar fields in cosmology is to consider them as classical, and to study their evolution. However, it is known that quantum fluctuations could modify the potential.\\
In this paper, we study the alterations of real or complex scalar field potentials due to quantum corrections. We focus on the one-loop quantum corrections to the original potentials, while computing effective potentials. We will therefore be able to test the validity of several dark energy, dark matter and dark fluid potentials. A similar study has been performed on several quintessence real scalar field models \cite{doran2002}, and we extend here that study to a broader range of cosmological models containing complex scalar fields or two coupled real scalar fields.\\
\\
The cosmological action is written as:
\begin{equation}
\mathcal{S} = \int \sqrt{|g|} \, d^4x \, \mathcal{L} \;\;,
\end{equation}
where $\mathcal{L}$ is the associated Lagrangian density. We consider that the Lagrangian contains a curvature term leading to Einstein equations, a term involving scalar fields minimally coupled to gravity, and a term containing fermions potentially coupled to the scalar fields \cite{doran2002}. We have then
\begin{equation}
\mathcal{L} = \mathcal{L}_{curv}+\mathcal{L}_{scalar}+\mathcal{L}_{fermion} \;\;.
\end{equation}
The curvature Lagrangian is
\begin{equation}
\mathcal{L}_{curv}=M_P^2 R \;\;,
\end{equation}
where $M_P$ is the Planck mass and $R$ the curvature tensor. For a model containing a complex scalar field $\Phi$, the scalar field Lagrangian can be written as:
\begin{equation}
\mathcal{L}_{scalar}=g^{\mu\nu} \partial_\mu\Phi^\dagger(x) \partial_\nu\Phi(x)-V(\Phi(x)) \;\;,
\end{equation}
where $V$ is the potential of the field. For two real scalar fields $\phi_1$ and $\phi_2$, the scalar field Lagrangian writes:
\begin{equation}
\mathcal{L}_{scalar}=\frac{1}{2} g^{\mu\nu} [\partial_\mu\phi_1(x) \partial_\nu\phi_1(x) + \partial_\mu\phi_2(x) \partial_\nu\phi_2(x)] -V(\phi_1(x),\phi_2(x)) \;\;.
\end{equation}
These two cases are completely equivalent, and the complex scalar field can be written in function of the two real scalar fields:
\begin{equation}
\Phi = \frac{1}{\sqrt{2}}(\phi_1+i\phi_2) \;\;.
\end{equation}
In presence of a single fermionic species coupled to the scalar fields, the fermionic Lagrangian reads:
\begin{equation}
\mathcal{L}_{fermion}=\bar{\Psi}^\dagger(x) [i \gamma^\mu \nabla_\mu - \gamma^5 m_f(\Phi)] \Psi(x)\;\;,
\end{equation}
where $m_f(\Phi)$ is the fermion mass, which is dependent on $\Phi$, $\gamma$'s are Dirac matrices and $\nabla_\mu$ is the covariant derivative.\\
We first restrict ourselves to the flat Minkowski space, so that $g^{\mu\nu}\rightarrow\eta^{\mu\nu}$, $\sqrt{|g|}\rightarrow 1$, $R\rightarrow 0$ and $\nabla_\mu\rightarrow\partial_\mu$.\\
The method we use here to study the quantum corrections is the saddle point expansion described in \cite{peskin,chengli}. The effective action of the scalar field writes, to higher order:
\begin{equation}
\Gamma[\Phi_{cl}]= \int d^4x \, \mathcal{L}_1[\Phi_{cl}] + \frac{i}{2} \log \det\left[\frac{\delta^2 \mathcal{L}_1}{\delta \phi_i \, \delta\phi_j}\right] + \cdots \;\;,
\end{equation}
where $\mathcal{L}_1$ is the renormalized part of $\mathcal{L}_{scalar}$, the subscript $cl$ indicates that the fields are classical, and $i,j=1,2$. The effective potential can be written in function of $\Gamma[\Phi_{cl}]$ such that
\begin{equation}
V_{\mbox{\small eff}}(\Phi_{cl})=-\frac{1}{VT}\Gamma[\Phi_{cl}] \;\;,
\end{equation}
where $VT$ is a space-time volume. We consider here only the one-loop corrections to the potential as shown in Fig~\ref{fig}. %
We need to define the effective masses
\begin{equation}
m^2_{ij}=m^2_{ji}=\frac{\partial^2 V}{\partial \phi_i \partial \phi_j} \;\;,
\end{equation}
and
\begin{eqnarray}
m_a^2 &=& \frac{1}{2} \left(m_{11}^2 + m_{22}^2 + \sqrt{(m_{11}^2 - m_{22}^2)^2 + 4 m_{12}^4}\right)\;\;,\\
m_b^2 &=& \frac{1}{2} \left(m_{11}^2 + m_{22}^2 - \sqrt{(m_{11}^2 - m_{22}^2)^2 + 4 m_{12}^4}\right) \;\;.
\end{eqnarray}
With a large momentum cutoff $\Lambda$, and ignoring the $\Phi$-independent contributions and the graphs of higher orders, the calculation of the effective potential due to the leading order scalar loop for two real coupled scalar fields leads to: 
\begin{equation}
\label{Vsca_tot}
V_{eff}(\Phi_{cl})=V(\Phi_{cl})+\frac{\Lambda^2}{32 \pi^2}(m_a^2 + m_b^2) +\frac{m_a^4 }{32\pi^2} \left[\ln\left(\frac{m_a^2}{\Lambda}\right)-\frac{3}{2}\right] + \frac{m_b^4 }{32\pi^2} \left[\ln\left(\frac{m_b^2}{\Lambda}\right)-\frac{3}{2}\right]\;\;.
\end{equation}
If the terms proportional to $m_a^4$ and $m_b^4$ are important in the context of usual field theory, in the case of cosmological scalar fields the potential is of the order of $10^{-123}$ $M_P^4$, and therefore we can safely disregard these terms in comparison to the terms proportional to $m_a^2$ and $m_b^2$. For this reason, even if many of the cosmological potentials are non-renormalizable in the strict sense of field theory, we can consider that the cosmological potentials are renormalizable as the higher order terms are so small. Therefore, the effective potential reduces to:
\begin{equation}
V_{1-loop}(\Phi_{cl})=V(\Phi_{cl})+\frac{\Lambda^2}{32 \pi^2} \left(\frac{\partial^2V}{\partial\phi_1^2}(\phi_{1\,cl},\phi_{2\,cl})+\frac{\partial^2V}{\partial\phi_2^2}(\phi_{1\,cl},\phi_{2\,cl})\right) \;\;.
\end{equation}
We can also calculate the correction due to fermions:
\begin{equation}
\label{Vfer_tot}
\delta V_{fermion}(\Phi_{cl})=-\frac{\Lambda_{ferm}^2}{8 \pi^2}[m_f(\Phi_{cl})]^2 - \frac{[m_f(\Phi_{cl})]^4 }{32\pi^2} \left[\ln\left(\frac{[m_f(\Phi_{cl})]^2}{\Lambda_{ferm}}\right)-\frac{3}{2}\right] \;\;.
\end{equation}
Again, because the fermion mass is very small in comparison to the Planck mass, the terms proportional to $[m_f(\Phi_{cl})]^4$ can be neglected. We obtain finally the one-loop effective potential for two real coupled scalar fields:
\begin{equation}
V_{1-loop}(\phi_{1\,cl},\phi_{2\,cl})=V(\phi_{1\,cl},\phi_{2\,cl})+\frac{\Lambda^2}{32 \pi^2} \left(\frac{\partial^2V}{\partial\phi_1^2}(\phi_{1\,cl},\phi_{2\,cl})+\frac{\partial^2V}{\partial\phi_2^2}(\phi_{1\,cl},\phi_{2\,cl})\right)-\frac{\Lambda_{f}^2}{8 \pi^2}[m_f(\phi_{1\,cl},\phi_{2\,cl})]^2 \;\;,
\end{equation}
or, for a single complex scalar field:
\begin{equation}
V_{1-loop}(\Phi_{cl})=V(\Phi_{cl})+\frac{\Lambda^2}{64 \pi^2} \frac{\partial^2V}{\partial\Phi^\dagger \partial\Phi}(\Phi_{cl})-\frac{\Lambda_{f}^2}{8 \pi^2}[m_f(\Phi_{cl})]^2 \;\;.
\end{equation}
These results are in agreement with the single real scalar field results of \cite{doran2002}.\\
We disregarded here the higher order terms, which is a reasonable approximation, as the potential and its derivatives are today of the order of the critical density, {\it i.e.} of the order of $10^{-123} M_P^4$. We also note that the ignored $\Phi$-independent contributions would lead to a cosmological constant of the order $\Lambda^4=\mathcal{O}(M_P^4)$, which is by far much larger than the critical density. This well-known problem appearing in the majority of field theories could hopefully be solved by some mechanism which could make these contributions vanish.\\
In the following, we will consider that $\Lambda = M_P$ unless stipulated otherwise, and we use units in which $c = \hbar = M_P = 1$. We study first the case in which the scalar fields are not coupled to fermions, for quintessence, scalar field dark matter and scalar field dark fluid models, and then we consider the case with a coupling to the fermions.
%
\section{Scalar fields without fermion coupling}
\noindent We calculate in this section the one-loop corrections to several cosmological scalar field potentials. First, we consider a single real scalar field associated to the SUGRA potential as in a dark energy quintessence model \cite{brax1999}. Then, we continue with the case of two coupled real scalar fields, involved in the coupled quintessence model \cite{bento2001}. We also 
discuss a dark matter model based on a complex scalar field associated to a quadratic \cite{arbey_quadratic} or quartic \cite{arbey_quartic} potential. Finally, we analyse the case of a unifying dark fluid model involving a complex scalar field associated to a potential containing a quadratic term and a decreasing exponential term \cite{arbey_darkfluid2}.
\subsection{SUGRA quintessence: single real scalar field}
\noindent The SUGRA potential for a quintessence real scalar field reads \cite{brax2006}:
\begin{equation}
V(\phi)=A \phi^{-\alpha}\exp(\beta \phi^2) \;\;,
\end{equation}
where $\beta = 8\pi / M_P^2$. We obtain the effective one-loop potential:
\begin{equation}
V_{1-loop}=V(\phi_{cl}) \beta^2 \frac{\Lambda^2}{8\pi^2}\phi_{cl}^2  \left[1 + \frac{1}{2 \beta^2}\left\{\frac{16\pi^2}{\Lambda^2} - (2\alpha-1)\beta\right\}\phi_{cl}^{-2} + \frac{1}{4 \beta^2}\alpha(\alpha+1)\phi_{cl}^{-4}\right] \;\;.
\end{equation}
It can be noted that the shape of the potential is changed. However, the global $\beta^2 \Lambda^2 / (8\pi^2)$ factor can be reabsorbed into the $A$ factor: $A \longrightarrow A (8\pi^2)/(\beta^2 \Lambda^2)$, and the $\phi_{cl}^2$ factor can also disappear with the redefinition $\alpha  \longrightarrow \alpha+2$. Thus, the shape can remain similar if \begin{equation}
\frac{1}{2 \beta^2}\left\{\frac{16\pi^2}{\Lambda^2} - (2\alpha-1)\beta\right\}\phi_{cl}^{-2} + \frac{1}{4 \beta^2}\alpha(\alpha+1)\phi_{cl}^{-4} \ll 1 \;\;.
\end{equation}
For a usual $\alpha = \mathcal{O}(10)$, as the present value of the field is of the order of $M_P$, this inequality is respected today. However, as already noticed in \cite{doran2002} concerning inverse power law potentials, problems could arise earlier as the correction can strongly modify the form of the potential. Fortunately, the argument developed in \cite{brax1999} should still hold for this kind of potential: the corrections are important only when quintessence was subdominant, and as corrections contains only negative power of $\phi_{cl}$, the field should be able to roll down the potential, and when the field becomes relevant to account for dark energy, the corrections may have become negligible. A more detailed analysis is nevertheless needed to check the validity of this scenario.
\subsection{Quintessence with coupled scalar fields: two real coupled scalar fields}
\noindent We consider now the potential for two coupled real scalar fields of \cite{bento2001}:
\begin{equation}
V(\phi_1,\phi_2)=\exp(-\lambda \phi_1) P(\phi_1,\phi_2) \;\;,
\end{equation}
with
\begin{equation}
P(\phi_1,\phi_2)=a+(\phi_1-\phi_1^0)^2+b(\phi_2-\phi_2^0)^2+c\,\phi_1(\phi_2-\phi_2^0)^2+d\,\phi_2(\phi_1-\phi_1^0)^2 \;\;.
\end{equation}
The derivatives of the potential reads:
\begin{equation}
\frac{\partial^2V}{\partial\phi_1^2}= \left[\lambda^2 P(\phi_1,\phi_2) - 2 \lambda \frac{\partial P}{\partial\phi_1} + \frac{\partial^2 P}{\partial\phi_1^2}\right]  \exp(-\lambda \phi_1) \;\;,
\end{equation}
and
\begin{equation}
\frac{\partial^2V}{\partial\phi_2^2}=(2 b + 2 c\, \phi_1) \exp(-\lambda \phi_1) \;\;,
\end{equation}
so that the effective potential becomes:
\begin{equation}
V_{1-loop}(\phi_{1\,cl},\phi_{2\,cl})= \exp(-\lambda \phi_{1\,cl}) P(\phi_{1\,cl},\phi_{2\,cl}) + \frac{\Lambda^2}{32 \pi^2} \left(\frac{\partial^2V}{\partial\phi_1^2}(\phi_{1\,cl},\phi_{2\,cl})+\frac{\partial^2V}{\partial\phi_2^2}(\phi_{1\,cl},\phi_{2\,cl})\right) \;\;.
\end{equation}
Using the same values for the parameters as in \cite{bento2001}, we can show that the additional terms which cannot be reabsorbed are presently of the same order as the value of the non-corrected potential. Therefore, the structure of the polynomial is globally changed, and unfortunately, the additional terms cannot be reabsorbed in the constants. A way out could be to replace $P(\phi_1,\phi_2)$ by a more general polynomial of the form
\begin{equation}
P(\phi_1,\phi_2)=\sum_{i+j=3} \alpha_{ij} \phi_1^i \phi_2^j \;\;,
\end{equation}
the $\alpha_{ij}$ being constants. In this case, the loop correction terms can be reabsorbed into the constants. Such a possibility has of course to be studied further, but would probably lack of credibility due to the high number of independent parameters.
\subsection{Dark matter complex scalar field}
\noindent We study also the following potential of a complex scalar field \cite{arbey_quadratic,arbey_cosmo,arbey_quartic}:
\begin{equation}
V(\Phi)=m^2 \Phi^\dagger \Phi + \lambda (\Phi^\dagger \Phi)^2 \;\;.
\end{equation}
The effective potential can be obtained:
\begin{equation}
V_{1-loop}= \frac{\Lambda^2}{64\pi^2}m^2 + \left(m^2+ \frac{\Lambda^2}{16\pi^2} \lambda \right) \Phi_{cl}^\dagger\Phi_{cl} +  \lambda (\Phi_{cl}^\dagger \Phi_{cl})^2 \;\;.
\end{equation}
As the effective potential is defined up to a constant, the constant term is irrelevant and can be safely disregarded. For $\lambda=0$, the potential is therefore unchanged. For $\lambda\neq0$, the potential is still stable, because the term $\Lambda^2/(16\pi^2) \lambda \Phi_{cl}^\dagger\Phi_{cl}$ can be reabsorbed into the redefinition $m^2 \longrightarrow m^2 - \Lambda^2/(16\pi^2) \lambda$. Thus, this potential is stable through loop corrections, and is in fact renormalizable. Of course, these results were predictable, as the quadratic and quartic potentials are well-known potentials which have been intensively studied.
\subsection{Dark fluid complex scalar field}
\noindent Finally, we consider the potential \cite{arbey_darkfluid2}:
\begin{equation}
V(\Phi)=m^2 \Phi^\dagger \Phi + \alpha \exp(-\beta \Phi^\dagger \Phi) \;\;.\label{DF1}
\end{equation}
This potential can lead to both dark matter and dark energy behaviors. The one-loop effective potential reads:
\begin{equation}
V_{1-loop}= \frac{\Lambda^2}{64\pi^2}m^2 + m^2 \Phi_{cl}^\dagger\Phi_{cl} + \frac{\Lambda^2}{64\pi^2}\alpha \beta^2 \ \Phi_{cl}^\dagger\Phi_{cl} \exp(-\beta \Phi_{cl}^\dagger\Phi_{cl}) + \alpha \left(1 - \beta \frac{\Lambda^2}{64\pi^2} \right) \exp(-\beta \Phi_{cl}^\dagger\Phi_{cl}) \;\;.
\end{equation}
Again, the constant term can be safely disregarded. The term $- \beta \Lambda^2/(64\pi^2) \exp(-\beta \Phi_{cl}^\dagger\Phi_{cl}) $ can be reabsorbed in the redefinition $\alpha \longrightarrow \alpha/\left(1 - \beta\Lambda^2/(64\pi^2) \right)$. However, a problem arises from the extra term $\Lambda^2/(64\pi^2) \alpha \beta^2 \Phi_{cl}^\dagger\Phi_{cl} \exp(-\beta \Phi_{cl}^\dagger\Phi_{cl})$ which modifies the potential. Considering the values of the parameters derived in \cite{arbey_darkfluid2}, this term cannot be neglected today and would lead to a strong modification of the potential, unless $\Lambda \ll 10^{-4}$. We can however note that at earlier epochs, as $|\Phi|$ was larger, this term was extincted by the exponential.\\
\\
A way to solve this problem would be to modify the potential such as:
\begin{equation}
V(\Phi)=m^2 \Phi^\dagger \Phi + (A + B \Phi^\dagger \Phi) \exp(-\beta \Phi^\dagger \Phi) \;\;.
\end{equation}
In this case, the effective potential writes:
\begin{eqnarray}
V_{1-loop} &=& \frac{\Lambda^2}{64\pi^2}m^2 + m^2 \Phi^\dagger \Phi  + \frac{\Lambda^2}{64\pi^2} B \beta^2 (\Phi_{cl}^\dagger\Phi_{cl})^2 \exp(-\beta \Phi_{cl}^\dagger\Phi_{cl})  \\ 
&& + \left[ \left(A + (B -  A \beta) \frac{\Lambda^2}{64\pi^2} \right) + \left(B + \frac{\Lambda^2}{64\pi^2} (A \beta^2 - 3 B \beta)\right) \Phi_{cl}^\dagger\Phi_{cl} \right] \exp(-\beta \Phi_{cl}^\dagger\Phi_{cl})\nonumber\;\;.
\end{eqnarray}
The constant term is irrelevant. With two redefinitions: $A \longrightarrow A - (B -  A \beta) \Lambda^2/(64\pi^2)$ and $B \longrightarrow B - (A \beta^2 - 3 B \beta) \Lambda^2/(64\pi^2) $, the correction terms of the second line vanish, but the last correction term of the first line still remains and modifies the potential shape. However, for $B \sim 10^{-120}$, this term is negligible in comparison to the other terms, and the shape of the potential is then conserved. If $A=\alpha$ and $B$ is small in comparison to $m^2$, the analysis of \cite{arbey_darkfluid2} would remain valid. In the case of a larger $B$, we could modify again the potential, such as:
\begin{equation}
V(\Phi)=m^2 \Phi^\dagger \Phi + \sum_n \bigl(\alpha_n (\Phi^\dagger \Phi)^n \bigr)\exp(-\beta \Phi^\dagger \Phi) \;\;, \label{DF2}
\end{equation}
where $n$ goes from 0 to $n_{max}$. In this case, the corrections can be reabsorbed into the constants $\alpha_n$, and the last correction of order $(\Phi_{cl}^\dagger \Phi_{cl})^{n_{max}+1}$ will hopefully be negligible today in comparison to the other terms. Yet, such a potential would lead to a behavior similar to that described in \cite{arbey_darkfluid2}, provided the redefined terms $\alpha_n (\Phi_{cl}^\dagger \Phi_{cl})^n$ (for $n \ne 0$) are today small in comparison to $\alpha_0$.\\
\\
To conclude this section, we have studied the one-loop corrections to several scalar field potentials. Some potentials can have difficulties to resist the quantum fluctuations. However, it is possible to build more suitable and robust potentials, in particular for the dark fluid model. 
%
\section{Scalar fields coupled to fermions}
\noindent We will now consider the corrections due to a coupling of the scalar fields to fermions. The fermion term in the effective potential reads:
\begin{equation}
V_{1-loop}^{f}(\Phi_{cl})=-\frac{\Lambda_{f}^2}{8 \pi^2}[m_f(\Phi_{cl})]^2 \;\;.
\end{equation}
If the fermion mass is $\Phi$-independent and of the order of 100 GeV, and considering that the fermionic cutoff is taken at GUT scale: $\Lambda_{f}=10^{-3}$, the fermionic term is of the order of $10^{-42}$, which is extremely large in comparison to the present value of the potential $V(\Phi_{cl}) \sim 10^{-123}$. Thus, as noticed in \cite{doran2002}, the fermionic corrections do not dominate the potential only if $[m_f(\Phi_{cl})]^2$ takes a form which is already contained in the classical potential, and which can then be reabsorbed in the constants of the potential.\\
\\
Let us focus for example on the dark fluid potential of Eq. (\ref{DF2}). The fermion mass can then take a form such as:
\begin{equation}
m_f(\Phi_{cl})=m_f^0 (\Phi_{cl}^\dagger \Phi_{cl})^{n/2} \exp\left(-\frac{1}{2} \beta \Phi_{cl}^\dagger\Phi_{cl}\right) \;\;,
\end{equation}
or the square root of a sum of similar terms. However, a simpler acceptable form would be:
\begin{equation}
[m_f(\Phi_{cl})]^2=(m_f^0)^2 \Phi_{cl}^\dagger \Phi_{cl} \;\;.
\end{equation}
In this case, $(m_f^0)^2$ can be reabsorbed by the redefinition $m^2 \longrightarrow m^2 + (m_f^0)^2$.\\
\\
We now consider the case where the coupling does not share the shape of the potential, and where the one-loop corrections can not be reabsorbed by higher order corrections. We can write the fermion mass as
\begin{equation}
m_f(\Phi_{cl})= m_f^0 + \delta m_f(\Phi_{cl}) \;\;.
\end{equation}
Assuming here that $\delta m_f(\Phi_{cl}) \ll m_f^0$, we have:
\begin{equation}
V_{1-loop}^{f}(\Phi_{cl})=-\frac{\Lambda_{f}^2}{8 \pi^2}[(m_f^0)^2+ 2 m_f^0 \delta m_f(\Phi_{cl}) + \delta m_f(\Phi_{cl})^2] \;\;.
\end{equation}
The constant term is pointless, as the effective potential is defined up to a constant. Thus, the global potential has a similar structure if
\begin{equation}
\frac{\Lambda_{f}^2}{4 \pi^2} m_f^0 \delta m_f(\Phi_{cl}) \ll V(\Phi_{cl}) \;\;,
\end{equation}
For a present value of the potential such as $V(\Phi_{cl}) \sim 10^{-123}$, this leads to a limit on the present value of $\delta m_f$:
\begin{equation}
\delta m_f(\Phi_{cl}) \ll 10^{-79} \mbox{ GeV} \;\;.
\end{equation}
This limit is so stringent that it nearly forbids a dependence of the mass on $\Phi_{cl}$ if the shape of the coupling is different from the shape of the potential. This result is similar for the quintessence potentials, and is in agreement with \cite{doran2002}. We however have to notice that as we restricted ourselves to one-loop corrections, this limit has to be interpreted with reserve, as the higher order corrections could also modify the structure of the effective potential.
%
\section{Other corrections}
\noindent We have just considered the effective one-loop quantum corrections to the scalar field potential. However, other corrections can also have an important influence on the scalar field behavior. In this section, we study the corrections due to the space-time curvature, and the corrections to the kinetic term.
\subsection{Curvature corrections}
\noindent To simplify, we discuss here the case of a single real scalar field. To study the effects of the curvature on the potential, $R \ne 0$ is assumed. We consider the same Lagrangian densities as in the introduction. Using the asympotic expansion of the Heat kernel for the involved operators, one can show that the corrections involving the curvature tensor which modify the dynamics of the scalar field read \cite{brax1999}:
\begin{eqnarray}
\delta V_{curv}(\Phi_{cl}) &=& \frac{1}{32\pi^2} \left( m_\Phi^2 - \frac{R}{6} \right)^2 \left[\ln\left(\frac{m_\Phi^2 - R/6}{\Lambda}\right)-\frac{3}{2}\right]\\
\nonumber &&- \frac{1}{32\pi^2} \left( [m_f(\Phi_{cl})]^2 + \frac{R}{12} \right)^2\left[\ln\left(\frac{[m_f(\Phi_{cl})]^2+R/12}{\Lambda_{ferm}}\right)-\frac{3}{2}\right] \;\;,
\end{eqnarray}
where
\begin{equation}
m_\Phi^2 = \frac{d^2 V}{d\Phi^2} \;\;.
\end{equation}
We can first note that in the case of a Minkowskian space $R=0$, the logarithmic terms of equations (\ref{Vsca_tot}) and (\ref{Vfer_tot}) are retrieved. Practically, the other correction terms are not modified by the curvature, and only the logarithmic terms that we disregarded receive contributions. Therefore, for small curvature, {\it i.e.} $R$ of the order of $m_\Phi^2$ or $m_f^2$, the terms which are involved here have a negligible impact on the effective potential at cosmological scales, and can be again neglected. The conclusion can be generalized to the case of two real or one complex scalar fields. However, more attention is of course needed in the case of a larger curvature.
\subsection{Kinetic term corrections}
\noindent The analyzes of the previous sections are valid provided the K\"ahler potential is flat. At low energy, under a scale $\Lambda_K$, the K\"ahler potential may be a more complicated function, called $K$, which can be expanded in Taylor series. For a single real scalar field, it writes:
\begin{equation}
\label{kahler_potential}
K(\Phi;\Lambda_K)= \Phi^2 + \sum_{n>0} a_n \frac{\Phi^{n+2}}{\Lambda_K^n}\;\;,
\end{equation}
where $a_n$ are constants. If $\Phi > \Lambda_K$, the K\"ahler potential is considered as flat, and therefore this expansion holds if $\Phi < \Lambda_K$.\\
In a flat space-time, the scalar field low-energy Lagrangian is
\begin{equation}
\mathcal{L}= \frac{1}{2} g \partial \Phi \partial \Phi - V(\Phi) \;\;,
\end{equation} 
where $g$ is the K\"ahler metric on the one dimensional curve defined as:
\begin{equation}
g = \frac{1}{2} \partial^2 K \;\;.
\end{equation}
To better render the physical meaning of the K\"ahler potential, we redefine the field such as
\begin{equation}
\frac{d\tilde{\Phi}}{d\Phi}= \sqrt{g} \;\;.
\end{equation}
Integrating and using an iterative inversion method, we can show that
\begin{equation}
\Phi = \tilde{\Phi} + \sum_{n>0} b_n \frac{\tilde{\Phi}^{n+1}}{\Lambda_K^n} \;\;,
\end{equation}
where the $b_n$ are constants which can be determined iteratively. With this redefinition, the kinetic term of $\Phi$ is transformed into a standard kinetic term for $\tilde{\Phi}$, and the potential writes:
\begin{equation}
V(\Phi) = V(\tilde{\Phi}) + \sum_{m,n>0} d_{mn} \frac{\tilde{\Phi}^{m+n}}{\Lambda_K^n} V^{(m)}(\tilde{\Phi}) \;\;,
\end{equation}
where $V^{(m)}$ denotes the $m$-th derivative of the potential $V$, and $d_{mn}$ are calculable constants.\\
Thus, the additional terms to the effective potential due to the kinetic term corrections are:
\begin{equation}
\delta V_{eff} = \sum_{m,n>0} d_{mn} \frac{\Phi^{m+n}}{\Lambda_K^n} V^{(m)}(\Phi) \;\;.
\end{equation}
We can first recall that the corrections appear if $\Phi < \Lambda_K$. Therefore, the energy scale $\Lambda_K$ has an important influence here. For our study, we can assume that $\Lambda_K$ is of the order of the Planck mass. The correction terms appear as powers of the scalar field times derivative of the potential. For the potentials we studied, because they involve either polynomials or exponential terms, they will keep their initial structure, but receive corrections under the form of powers of the scalar fields.\\
For the dark matter and dark fluid potentials, as the kinetic corrections are small, they should not perturb much the dynamics of the scalar fields, at most providing some small long-range effects. However, in the case of quintessence models, a more detailed study would be needed to determine if these corrections could perturb the tracking properties, and then the behavior the fields.\\
To conclude this sections, we have seen that other corrections can perturb the scalar fields. However, in models with  small curvature or nearly flat K\"ahler potential, the considered corrections can be neglected. 
%
\section{Conclusion}
\noindent In this letter, we performed the calculation of the one-loop effective potentials of several cosmological models based on scalar fields. We have seen that some potentials are stable against the quantum fluctuations, and in particular the simple quadratic and quartic potentials of the dark matter scalar field models. Different solutions to the instabilities of the coupled scalar field quintessence potential and of the dark fluid potential have been proposed.\\
Focusing on the fermionic corrections, we have confirmed the results of \cite{doran2002} with regard to the extreme restriction of the coupling to the fermions. This constraint is relaxed if the coupling has a common structure with the scalar field potential or can be compensated by higher order corrections.\\
Concerning corrections due to the curvature or the kinetic term, we have seen that in usual case they are small, but they may have to be carefully studied in the case of strong curvature of the space-time or the K\"ahler potential.\\
The calculations we performed in this work concern only the one-loop order corrections, and therefore do not give access to the true effective potential. However, such studies reveal to be valuable tools to test and build robust scalar field potentials.
%
\section*{Acknowledgements}
\noindent A.A.'s research is partially funded by a BQR grant from Universit\'e Lyon~1, and he acknowledges the support from the Horizon Project as well. F.M. acknowledges the support from NSERC.
%

\end{document}